# Using Molecular Dynamics to Model the Binding Affinities of Nitrogen-Containing Bisphosphonates on Hydroxyapatite Surface

Tzu-Jen Lin*

Department of Chemical Engineering, Chung Yuan Christian University, 200 Chung Pei Road, Chung Li District, Taoyuan City, Taiwan 32023

**Abstract**

Nitrogen-containing bisphosphonates (NBPs) are widely used to treat bone density loss and skeletal disorders. The adsorption, retention, diffusion, and release of (NBPs) in bone minerals are governed by their binding affinities to such minerals. Extensive crystal growth, nuclear magnetic resonance (NMR), and isothermal titration calorimetry (ITC) studies have been conducted to measure the binding affinities of NBPs to bone minerals, but the results have been inconsistent. In this work, we examined the binding free energies of zoledronate (ZOL), risedronate (RIS), pamidronate (PAM), alendronate (ALN), and ibandronate (IBN) to neat and protonated hydroxyapatite (HAP) surfaces including (001), (010), and (101) facets through molecular dynamics. Our simulation results showed that nitrogen-containing functional groups and surface protonation influence the binding affinities of BPs to HAP surfaces. The hydrophilic behavior of nitrogen-containing functional groups enabled ZOL have a larger binding free energy than RIS and IBN at neat HAP surfaces. The protonated and slender amine groups of PAM and IBN enabled them have larger binding free energies and wider potential wells than that of ZOL, RIS, and IBN at neat and protonated HAP surfaces. In general, surface protonation in general reduced the binding affinities of NBPs except protonated (010) surfaces. NBPs preferred to immerse at the grooves of protonated (010) surfaces with high binding free energies, and the immersions were consistent with a site-binding model developed from (ITC) measurements. Based on our simulations results we propose that the binding rank is ALN > PAM > ZOL > IBN > RIS which is consistent with NMR and ITC studies, and the rank at neat (001) surfaces is ZOL > ALN > IBN > RIS, which correlates with crystal growth studies. The simulation results helped to unify diverged observations in the literature.

# 1. Introduction

Bisphosphonates (BPs) are common medications for treating osteoporosis and other skeletal disorders because of their high binding affinities to bone minerals composed majorly of calcium phosphates. They are also good candidates for bone-targeted drug delivery and diagnostic imaging.[1, 2] Pharmacokinetic studies have reported that 50% of injected BPs were absorbed by the skeleton and remained within the bones,[3] and that osteoporosis treatment relies on the long-term binding of BPs to bone minerals. The selective binding to bone minerals also facilitates osteoclast inhabitation and differentiation during bone resorption.[4-6] Moreover, the binding affinities of BPs can influence their distribution, diffusion, and release in bones. Therefore, it is crucial to understand how binding affinity of BPs to bone minerals is influenced by their molecular structures.

Figure 1 presents generic structure of BPs, which is characterized by two phosphonate groups bonded to the same carbon atom (P-C-P link). Moreover, substitutions of the P-C-P link can occur at two positions, R1 and R2. Hydroxyapatite (HAP)[7, 8] which is the most stable phase of calcium phosphates, is usually used to represent bond minerals in experiments. Experimental studies have suggested that two phosphonate groups in BPs are required for high binding affinities to HAP surfaces. Furthermore, a hydroxyl group at the R1 position improved the binding to bone mineral surfaces through the coordination of the hydroxyl group and calcium atoms at HAP surfaces.[9, 10] In the last decade, experimental studies have focused on BPs with various nitrogen-containing functional groups at the R2 position because of their high binding affinities to bone mineral surfaces and inhibition to bone resorption. However, the binding affinity rank of nitrogen-containing bisphosphonates (NBPs) to bone mineral surfaces has not reached conclusions yet. Nancollas and colleagues used kinetic studies on the crystal growth of HAP to determine the binding affinities of NBPs to HAP surfaces with the rank order zoledronate (ZOL) > pamidronate (PAM) > alendronate (ALN) > ibandronate (IBN) > risedronate (RIS) > etidronate > clodronate.[9] The rank order was the same as that determined by Henneman and colleagues who used dissolution studies on carbonated apatites.[11] Russel and colleagues suggested that ZOL has large binding affinities to bone mineral surfaces, because of its long retention time in HAP chromatography measurements.[12] Using a nuclear magnetic resonance (NMR)-based in vitro assay, Jahnke and colleagues suggested that the binding affinity rank was PAM> ALN > ZOL >

RIS > IBN in HAP and ALN > PAM > ZOL > RIS > IBN in bone powder.[13] Mukherjee and colleagues found the same rank order as Jahnke did.[14,15] Moreover, Leu and colleagues suggested that clinical BPs have similar affinities to human bone.[16]   Other than binding affinity rank to bone mineral surfaces, the binding mechanism of NBPs remains an open question. The binding of NBPs to HAP surfaces has been attributed to their phosphonate groups interacting with surface calcium atoms or nitrogen-containing functional groups forming N-H-O hydrogen bonds on surfaces.[10,17] On the other hand, Mukherjee and colleagues used NMR to propose an alternative site-binding model, and stated that NBPs bind to HAP surfaces by displacing surface phosphate ions.[14,15]

Molecular simulations are useful for studying interactions between organic molecules and biomineral surfaces.[18-24] HAP has also been used to represent bone minerals in molecular simulations. However, several issues have made current molecular simulation results of NBPs binding to HAP surfaces difficult to correlate with experimental measurements. The first issue is the inclusion of water molecules in simulation models. Robinson and colleagues used molecular mechanics to compute the binding energies of BPs to HAP surfaces, but their model did not include water molecules.[25] Ri et al. and Canepa et al. used ab initio methods to study the binding energies of zoledronic and alendronic acid to HAP surfaces, respectively, and their studies did not include water molecules.[26, 27] Chen and colleagues used molecular dynamics to compute the binding energies of BPs at HAP-water interfaces, but their model considered water layers close to HAP surfaces.[28] Most binding affinity measurements of BPs to HAP surfaces have been conducted in an aqueous environment, so molecular simulation models should include sufficient water molecules. The second issue is force field and models for HAP surface. Universal Force Field (UFF), Generalized Assisted Model Building with Energy Refinement Force Field (GAFF), or refitted Born-Mayer-Huggins potential was implemented for HAP surface models in literature.[18, 25, 28-30] Nevertheless, these force fields have not been validated with respect to the thermodynamic properties at HAP-water interfaces. Additionally, most molecular simulations have considered only neat-cleaved HAP surfaces. In fact, HAP surfaces undergo protonation in an aqueous environment, and NMR, infrared spectroscopy (IR), and atomic force microscopy (AFM) studies have observed protonated phosphates at HAP surfaces in an aqueous environment.[31-33] The third issue is that simulations have calculated the binding energies rather than binding free energies of NBPs to HAP surfaces.[25, 28] This has made it difficult for simulation results to correlate with binding affinity measurements in experiments because binding affinity relates to binding free energy rather than binding energy.

In this study, we used validated force field and surface models of HAP to study binding free energies and binding mechanisms of selected NBPs, ZOL, RIS, PAM, ALN, and IBN, at neat and protonated HAP surfaces through umbrella sampling methods. The computed binding free energies of selected NBPs were compared with experimental results, and the binding mechanism was examined.

## 2. Model and Simulation Method

### 2.1 Nitrogen-Containing Bisphosphonate Model

The molecular structure of ZOL, RIS, IBN, PAM, and ALN are presented in Figure 1. We considered the protonation form of IBN, PAM, and ALN because the pKa values of their side chain amine groups are larger than 7. In this study, protonated IBN, PAM, and ALN are referred to as IBNP, PAMP, and ALNP respectively. The pKa value of imidazole group in ZOL is close to 7, which means the ratio of protonated to neutral ZOL is 1:1 in a pH 7 condition. Therefore, we considered both the protonation and neutral forms of ZOL, which are referred to as ZOLP and ZOL respectively. Because the pKa value of pyridine group in RIS is much lower than 7, we consider its neutral state. GAFF[34] was applied to NBPs. Their atomic charges were obtained through an ab initio method. The molecular structures of the NBPs were using M06-2X functional with def2-SVP basis sets and the Solvation Model based on Density (SMD).[35-37] The optimized structures were used for atomic charge assignments through HF/6-31g* calculations combined with restrained electrostatic potential (RESP) method.[38] The details of the atom types and atomic charges of the NBPs are presented in Tables A1 to A6 in Appendices.

### 2.2 HAP Surface Model

The INTERFACE force field of HAP was developed by Heinz et al. and Lin et al. The force field was not only parameterized for bulk properties but also for interfacial thermodynamic properties such as cleavage energy, solid-liquid interface energy, and immersion energy.[39, 40] Moreover, Lin and colleagues proposed a strategy for constructing protonated HAP surfaces under different pH environments.[40] In the current study, we applied this strategy to create HAP surfaces corresponding to a pH 7 environment.

Based on the pKa values of phosphate ions, we assumed that two-thirds of surface phosphates are double protonated ($H_2PO_4$) and one-third are single protonated ($HPO_4$) when HAP surfaces are in a pH 7 condition. This assumption was confirmed by Tanaka and colleagues through immersion energy measurements on synthetic HAP.[41] The protonation scheme of HAP surfaces were constrained by stoichiometric chemistry and charge neutrality by removing calcium and hydroxide ions and adjusting the atomic charges of the protonated phosphates at the surface.[40] The setups of the atomic charges of protonated phosphates at HAP surfaces are presented in Figure A1 in Appendices. Immersion energies were computed to validate protonated HAP surface models. Immersion energy is the heat released when a solid is immersed in a liquid. In other words, it is the surface energy changes from a solid-vacuum to a solid-liquid interface. Therefore, the immersion energy per unit of surface area was obtained by subtracting the energy of solid-vacuum interface and bulk water models from the energy of a solid-water interface model, divided by the surface dimension. All simulation boxes equilibrated in NPT ensemble at 1 atm and 298 °K, whereas the energy measurements were conducted in NVT ensemble at 298 °K. The computed immersion energies of HAP (001)-pH7, (010)-pH7, and (101)-pH7 surface models ranged from 500 to 700 mJ/m$^2$, which were in the range of experimental measurements by Barton and Harrison and as well as Tanaka and colleagues.[41, 42] Some measurements reported low immersion energies of HAP from 230 to 470 mJ/m$^2$, but we assumed that the samples had been hydrated before measurements.[43] The computed immersion energy revealed that HAP (010) facet favored water than the other two facet at pH 7 condition, indicating that the (010) facet was the dominate facet in aqueous environments. This is consistent with crystal-growth studies.[44-46] We assessed the binding free energy of NBPs on both neat and protonated HAP surfaces because we wanted to see how surface protonation influences the binding affinities of NBPs.

## 2.3 Binding Free Energy Calculation

HAP-water-BP interface models of neat and protonated surfaces were developed for binding free energy calculations. Each HAP-water-BP interface model contained one HAP surface slab, 1500 water molecules, one NBP, and counter ions. The periodic condition was implemented to x, y, and z directions. The heights of the surface slabs were 27 Å to 33 Å for different facets, and the surface areas were approximately 35 Å x 35 Å. SPC/E rigid water model was used in this study.[47] The HAP-water-BP interface models equilibrated in NPT ensemble at a standard condition, 1 atm and 298.15 °K, for 4 ns, and Parrinello-Rahman anisotropic barostat and Nose-Hoover chain thermostat were implemented to regulate the pressures and temperatures respectively.[48-50] The simulation boxes after NPT equilibration were used for further binding free energy calculations.

Calculating the binding free energy of NBPs was completed by umbrella sampling, which is a biased sampling method to obtain free energy differences between thermodynamic states.[51, 52] The methodology restricts a system to a set of windows along a pre-defined reaction coordinate between thermodynamic states. These restrictions were completed by adding biased potentials in each window that generated biased histograms (probability distributions). Next, the biased probability distribution of each window was collected and reweighted by using the Weighted Histogram Analysis Method (WHAM) to reconstruct free energy profiles along a pre-defined reaction coordinate.[53, 54]

The reaction coordinate in this study was the distance between the mass center of a NBP and the mass center of the HAP surface slab. The direction was normal to HAP surfaces. The reaction coordinate started from the mass center of NBPs located at surface oxygen atoms to the mass center located at 15 Å from surface oxygen atoms. Before umbrella sampling, the NBPs were placed at the center of HAP surfaces. Overall, 20 to 25 windows were constructed for each umbrella sampling along the reaction coordinate, and each window was separated by 0.3 Å to 1.0 Å. The force constants of biased harmonic potential were 5000 to 10,000 kJ/mol nm$^2$. All umbrella samplings were conducted in NVT ensemble at 298.15 °K and Nose-Hoover chain thermostat was used. The velocity-verlet algorithm was implemented to integrate Newton's equation of motion with a time step of 1.0 fs. LINCS algorithm was implemented to constrain the motion of covalent-bonded hydrogen atoms.[55] Each window was equilibrated 2 ns to 8 ns followed by 4 ns production runs to construct sufficient overlaps between histograms for the WHAM.[56] All simulation works in this study were completed in GROMACS 5.1, and VMD was used for visualizations.[57-59]

## 3. Results and Discussions

The computed binding free energies of selected NBPs to neat and protonated HAP surfaces are presented in Tables 1 and 2, respectively. The values of the binding free energies were identified by the most negative values of computed free energy profiles along reaction coordinates. The computed free energy profiles are presented in Figures A2 and A3 of Appendices. The two rightmost columns of Tables 1 and 2 also list the measured binding free energies of NBPs to bone mineral surfaces by using isothermal titration calorimetry (ITC). Mukherjee and colleagues [14] used one and two site-binding models to explain measured weak and strong binding values. The average weak and strong binding values of selected NBPs were -5.2 and -8.5 kcal/mol, respectively.

### 3.1 Zoledronate and Protonated Zoledronate

ZOL and ZOLP bound to neat HAP facets by their phosphonate parts which are presented in Figure 3. The phosphonate parts interacted with surface phosphates and calcium ions through hydrogen bonds and electrostatic interactions, respectively. ZOL and ZOLP showed similar binding free energies to all neat HAP surfaces between −4.0 and −7.0 kcal/mol. Generally, the binding free energy of ZOLP was 1.5 kcal/mol larger than that of ZOL. This was because the protonated imidazole ring of ZOLP is hydrophilic, which increased the binding stability in the water monolayer at HAP-water interfaces. The relative hydrophilic characters between ZOL and ZOLP were confirmed by solvation free energies. The solvation free energy of ZOLP (−59.6 kcal/mol) was 30 kcal/mol larger than that of ZOL (−27 kcal/mol). (see Table A7 of Appendices) When it came to pH 7 HAP surfaces, the binding free energy of ZOL and ZOLP reduced to less than −2.7 kcal/mol and showed non-binding behavior at (001)-pH7 and (101)-pH7 surfaces based on computed free energy profiles. Because the HAP surfaces were protonated, less calcium ions remained and the protonated phosphate ions became less negatively charged at surfaces. This reduced the binding affinities of ZOL and ZOLP to these two surfaces, and the results also suggested that their binding affinities were highly sensitive to the crystallinity of the surface structure. Conversely, ZOL and ZOLP showed substantial binding affinities to (010)-pH7 surfaces with the values −6.4 and −11.8 kcal/mol, respectively. We found that ZOL and ZOLP slipped into the grooves of (010)-pH7 surfaces. (see Figure 5)

The absence of calcium and hydroxide ions at (010)-pH7 surfaces from the protonation scheme made ZOL and ZOLP slip into the grooves easily, leading to direct interactions to calcium and phosphate ions at the bottom of surface layers. ITC measured two binding free energy values for ZOL: −5.4 and −8.4 kcal/mol.[14] Because the composition ratio of ZOL and ZOLP was 1:1 at pH 7, we averaged the computed binding free energy of ZOL and ZOLP (ZOL* in Tables 2 and 3). We found the binding free energy of ZOL* at all neat HAP surfaces and (010)-pH7 HAP surfaces correlated with weak and strong binding values of the ITC results, respectively. This suggested that the weak binding values corresponded to ZOL binding to the well-crystallized region of HAP surfaces and that the strong binding values corresponded to ZOL immersed in the grooves of (010)-pH7 surfaces.

## 3.2 Risedronate and Ibandronate

RIS and IBNP showed weak binding free energies and non-binding behaviors to neat HAP surfaces based on the computed binding free energy profiles. RIS and IBNP interacted with surface phosphates and calcium ions by their phosphonate parts, which were the same as ZOL and ZOLP (see Figure 3). Nevertheless, their binding free energies were less than 2 kcal/mol at neat HAP surfaces. Therefore, we proposed that the pyridine part of RIS and the aliphatic side chain of IBNP reduce their binding affinities to neat HAP surfaces in aqueous environments. Based on computed solvation free energies in Table S5, RIS showed lowest solvation free energy (−23.9 kcal/mol) among the other NBPs in this study. Therefore, the hydrophobic character of RIS decreased its binding stability at the monolayer at HAP-water interfaces. The aliphatic amine part of IBNP was protonated, but it was shielded by its nearby methyl and bulky aliphatic side chains. The shielded charge and aliphatic bulky groups of IBNP increased its hydrophobic character at HAP-water interfaces by reducing interactions to charged HAP surfaces. Regarding pH 7 surfaces, IBNP showed binding free energy, −5.4 kcal/mol, at (010)-pH7 surfaces, but RIS still showed non-binding behavior to pH 7 HAP surfaces. By summarizing the computed binding free energies of RIS and IBNP at neat and pH 7 HAP surfaces, we concluded that IBNP had a slightly larger binding affinity than RIS to HAP surfaces. This is qualitatively consistent with crystal growth studies by Nancollas and colleagues.[9] Moreover, RIS and IBNP had lower binding free energies than other BPs in this study, which is qualitatively consistent with NMR study by Jahnke. The ITC measurements showed weak and strong binding values of RIS −4.7 and −7.3 kcal/mol, respectively, which were different from the computed ones. Additionally, the facet average of the computed binding free energy of IBNP at protonated surfaces was −3.8 kcal/mol, which was slightly less than the average weak binding value of BPs from the ITC measurements (−5.2 kcal/mol). Our simulations qualitatively predicted the relative binding affinity of RIS and IBNP, however, quantitatively underestimated their binding free energies to HAP surfaces. This was probably because force field parameters of RIS and IBNP gave them too much hydrophobic behavior.

### 3.3 Protonated Pamidronate and Alendronate

PAMP and ALNP had different binding mechanisms than the other NBPs in this study. PAMP and ALNP anchored phosphate ions at neat HAP surfaces mainly by their charged amine groups. Moreover, when they were 4 Å to 5 Å from HAP surfaces, their aliphatic amine groups were at fully extended conformations and trying to interact with surfaces. Therefore, their free energy profiles showed wide potential wells, as presented in Figure A3 in Appendices. At neat (001) surfaces, PAMP and ALNP showed similar binding free energies (−5.5 kcal/mol) to ZOL and ZOLP. PAMP and ALNP had larger binding free energies, −6.4 and −9.6 kcal/mol, respectively, at neat (010) surfaces than at neat (001) surfaces. This was attributed to their slender aliphatic amine groups. PAMP and ALNP possessed fully extended conformation and their aliphatic amine groups were able to slip into the grooves of neat (010) surfaces leading to strong electrostatic interactions between their charged amine groups and phosphate ions beneath the surface (see Figure 4). Conversely, the bulky nitrogen-containing functional groups of ZOL, ZOLP, RIS, and IBNP were not able to slip into such grooves. The binding free energy of ALNP was 3 kcal/mol larger than that of PAMP. This was because at fully extended conformation, the longer aliphatic charged amine group of ALNP created larger dipole moment than that in PAMP. PAMP and ALNP had free energies at neat (101) surfaces, −10.2 and −17.2 kcal/mol, respectively, which were larger than their values at neat (001) and (010) surfaces. The large binding free energy was attributed to their charged amine and phosphonate groups, which both interacted with surface phosphates and calcium ions. ALNP also had a larger binding free energy than PAMP at neat (101) surfaces. This was because the longer aliphatic amine chain of ALNP gave more conformational flexibility than that in PAMP, enabling both parts to interact with surface ions completely. Conversely, we found that the phosphonate groups of PAMP partially interacted with surface ions (see Figure 4).

The binding free energies of PAMP and ALNP were −7 and −6 kcal/mol at (101)-pH7 surfaces and −4 and −6 kcal/mol at (001)-pH7surfaces, respectively. These values indicated that PAMP and ALNP still strongly bound to these two protonated HAP surfaces. Electrostatic interactions between their charged amine groups and protonated phosphate ions at surfaces were the reason why they maintained substantial bindings to these two surfaces. PAMP and ALNP slipped into the grooves of (010)-pH7 surfaces, as with other NBPs in this study. The absence of calcium and hydroxide ions at (010)-pH7 surfaces not only enabled PAMP and ALNP to slip into grooves of the surface easily but also reduced van der Waals repulsions. These two factors led to the high binding free energies of PAMP (−14.7 kcal/mol) and ALNP (−21.9 kcal/mol) at the surfaces. ALNP had a larger binding free energy than PAMP at (010)-pH7 surface. We suggest that the reason for this was the same as that proposed for neat (010) surface: the longer aliphatic side chain of ALNP created larger dipole moments than that in PAMP at fully extended conformations. The facet-averaged binding free energies of PAMP and ALNP at neat (PAMP: -7.6 kcal/mol, ALNP: −10.7 kcal/mol) and protonated (PAMP: -8.6 kcal/mol, ALNP: −11.3 kcal/mol) HAP surfaces were close to the strong binding values in the ITC measurements. The simulation results revealed that PAMP and ALNP had strong binding at either neat or protonated HAP surfaces. The weak binding values of PAMP and ALNP from the experiments may have been associated with their binding at (001)-pH7 and (101)-pH7 surfaces, and the strong binding values may have corresponded to many of PAMP and ALNP binding to (010)-pH7 surfaces in the ITC measurements

## 4. Conclusions

In this study, we examined how nitrogen-containing functional groups and surface protonation influenced the binding affinities of ZOL, RIS, IBN, PAM, and ALN through molecular simulation. Hydrophilic and hydrophobic behaviors of their nitrogen-containing functional groups differentiated their binding affinities to neat HAP surfaces. ZOLP had larger binding free energies than RIS and IBNP at neat HAP surfaces because protonated imidazole group of ZOLP were much more hydrophilic than the pyridine and bulky aliphatic groups of RIS and IBNP respectively. Computed binding free energies of ZOL and ZOLP to neat HAP surfaces matched the weak binding values in ITC measurements. RIS and IBNP showed the weakest binding free energies at neat and protonated HAP surfaces compared with the other NBPs in this work, which correlated with NMR studies qualitatively.

Generally, protonated aliphatic amine groups of PAMP and ALNP enabled them to have larger binding free energies to neat HAP surfaces than that of ZOL, ZOLP, RIS, and IBNP. This was because of strong electrostatic interactions between protonated amine groups and surface phosphate ions. Even if PAMP and ALNP was 4 Å to 5 Å from surfaces, their protonated amine groups still tried to interact with surfaces leading to wide potential wells in their binding free energy profiles. The slender and protonated amine groups of PAMP and ALNP enabled them to slip into the grooves of neat (010) surfaces. Moreover, the longer aliphatic charged amine group of ALNP led to larger dipole moment at fully extended conformation and more conformational flexibility than that of PAMP. These factors enabled ALNP to have a higher binding free energy than PAMP at neat HAP (010) and (101) surfaces.

Surface protonation reduced the binding affinities of NBPs to (001)-pH7 and (101)-pH7 surfaces because fewer calcium ions and less negatively charged protonated phosphates remained at the surface. The binding free energies and profiles of ZOL and ZOLP showed non-binding behavior at (001)-pH7 and (101)-pH7 surfaces, indicating that their binding was highly sensitive to crystallinity of HAP surfaces. PAMP and ALNP still maintained substantial bindings to these two surfaces. This was because of the strong electrostatic interactions between their protonated aliphatic amine groups to surface protonated phosphate. The absence of calcium and hydroxide ions at (010)-pH7 surfaces eased van der Waal repulsions to enable NBPs to immerse in the surface grooves easily leading to large binding free energies. Nevertheless, RIS and IBNP still showed weak binding to (010)-pH7 surfaces because of their hydrophobic behaviors. The immersion of NBPs at (010)-pH7 grooves was similar to the two-site-binding model proposed by Mukherjee and colleagues.[14,15]

Based on our simulation results, binding affinities of NBPs were as follows: ALN > PAM > ZOL > IBNP > RIS at neat and protonated HAP surfaces in general. This result is similar to the NMR study by Jahnke and colleagues and ITC measurements by Mukherjee and colleagues.[13-15] Moreover, the binding affinity order at neat (001) surfaces was consistent with the crystal growth study by Nancollas and colleagues: ZOL > ALN > IBN > RIS.[9] The simulation results united the diverged observations in experimental studies. Moreover, this study showed that the INTERFACE Force Field for HAP has great potential for qualitatively and quantitatively predicting interactions between organic molecules and HAP surfaces. We believe that combining proper simulation technique and force field parameter will be helpful to drug design for osteoporosis and drug targeting in human bones and teeth.

## 5. Acknowledgements

This research was supported by the Ministry of Science and Technology, Taiwan (105-2218-E-033 -010 -MY2). The author is grateful for the generous allocation of computational resources provided by the National Center for High-Performance Computing, Taiwan.

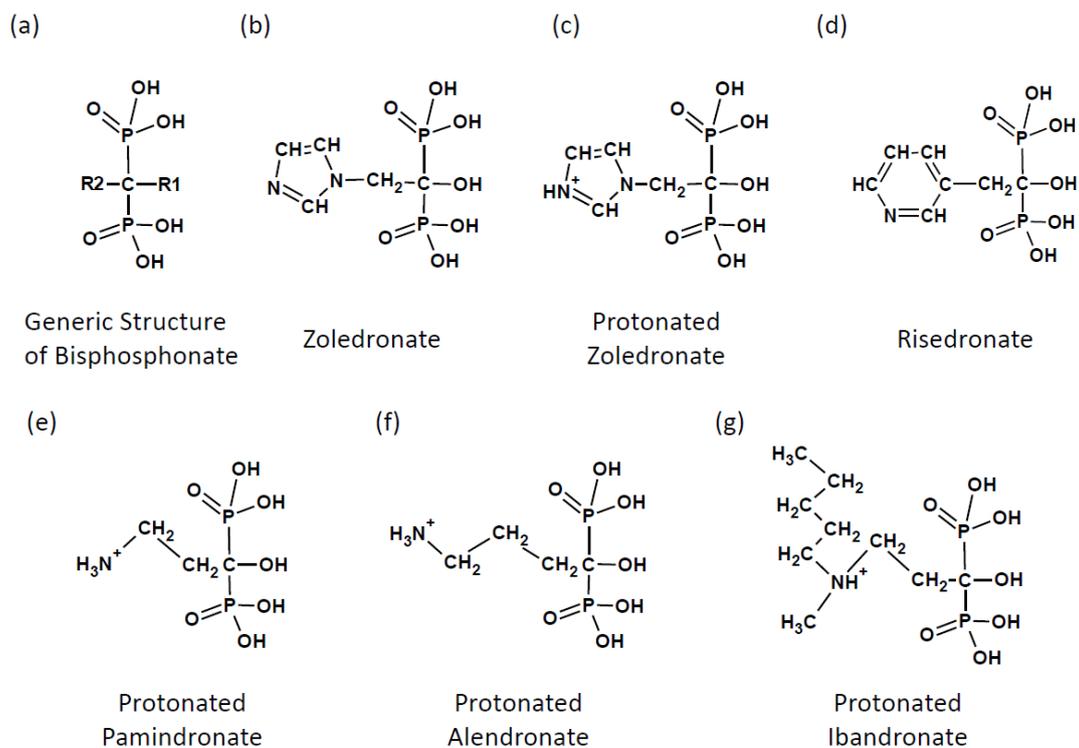

Figure 1. Molecular structures of (a) a generic bisphosphonate and nitrogen-containing bisphosphonate used in this study, (b) Zoledronate (ZOL), (c) Protonated Zoledronate (ZOLP), (d) Risedronate (RIS), (e) Protonated Pamidronate (PAMP), (f) Protonated Alendronate (ALNP), and (g) Protonated Ibandronate (IBNP).

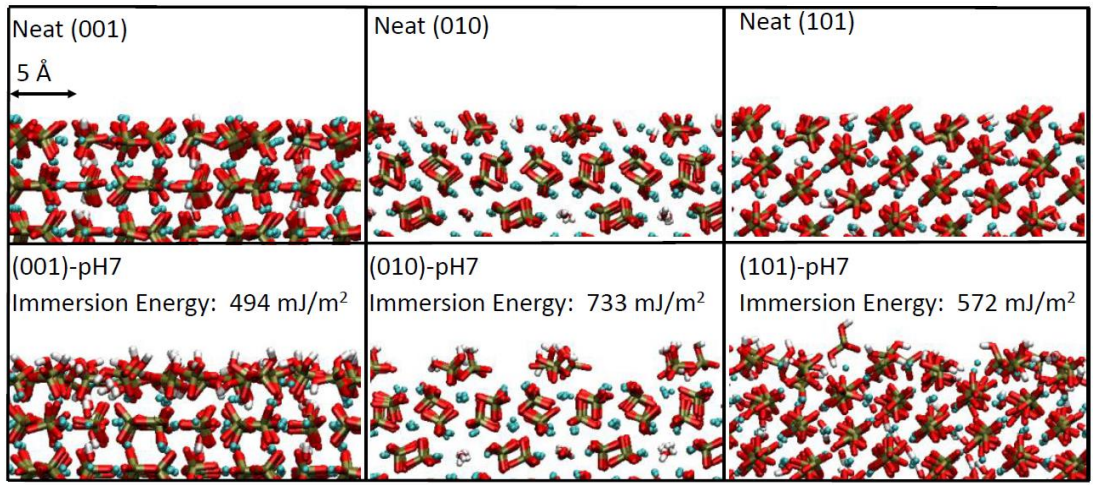

Figure 2. Side view of equilibrated HAP surface models in neat and pH 7 conditions. The surface structures in a pH 7 condition were more relaxed and disordered than the surface structures in neat condition because of the removal of superficial calcium and hydroxide ions reducing electrostatic interactions between ions at the surfaces. Additionally, the protonated phosphates in a pH 7 condition were less negative than the phosphates in the neat condition, also resulting in fewer electrostatic interactions. The computed immersion energies of each pH 7 surface are listed. The red, white, green, and brown atoms represent oxygen, hydrogen, calcium, and phosphorous atoms, respectively.

Table 1. Binding free energy of nitrogen-containing bisphosphonates at neat (001), (010), and (101) surfaces. The energy unit was kcal/mol. ZOL* represents the average binding free energies of ZOL and ZOLP. The two rightmost columns show the strong and weak binding values from the ITC measurements.[14]

| Neat Surface | (001) | (010) | (101) | ITC Weak | ITC Strong |
|---|---|---|---|---|---|
| ZOL | −5.5± 0.3 | −4.2± 0.4 | −5.1± 0.4 | −5.4 | −8.4 |
| ZOLP | −7.2 ± 0.6 | −5.7± 0.4 | −6.5± 0.4 | | |
| ZOL* | −6.4± 0.5 | −5.0± 0.4 | −5.8± 0.4 | | |
| RIS | −1.6± 0.5 | −0.8± 0.3 | −1.0± 0.3 | −4.7 | −7.3 |
| IBNP | −2.3±0.3 | −0.5±0.4 | −2.2± 0.3 | N/A | N/A |
| PAMP | −5.7± 0.4 | −6.4± 0.3 | −10.7 ± 0.5 | −5.7 | −9.3 |
| ALNP | −5.3± 0.4 | −9.6± 0.3 | −17.2± 0.2 | −6.5 | −10.4 |

Table 2. Binding free energy of nitrogen-containing bisphosphonates at protonated (001), (010), and (101) surfaces in a pH 7 condition. The energy unit was kcal/mol. ZOL* represents the average binding free energies of ZOL and ZOLP. The two rightmost columns show the strong and weak binding values from the ITC measurements.[14]

| pH7 Surface | (001) | (010) | (101) | ITC Weak | ITC Strong |
|---|---|---|---|---|---|
| ZOL | −1.6± 0.5 | −6.4± 0.2 | 0.0 ± 0.2 | −5.4 | −8.4 |
| ZOLP | −2.7 ± 0.3 | −11.8 ± 0.5 | −1.3 ± 0.2 | | |
| ZOL* | −2.2± 0.4 | −9.1± 0.4 | −0.7± 0.2 | | |
| RIS | −2.6± 0.2 | −1.4± 0.4 | −1.7 ± 0.2 | −4.7 | −7.3 |
| IBNP | −3.1± 0.5 | −5.4± 0.5 | −3.0± 0.4 | N/A | N/A |
| PAMP | −3.8± 0.1 | −14.7± 0.2 | −7.3 ± 0.2 | −5.7 | −9.3 |
| ALNP | −5.9± 0.4 | −21.9± 0.5 | −6.2± 0.4 | −6.5 | −10.4 |

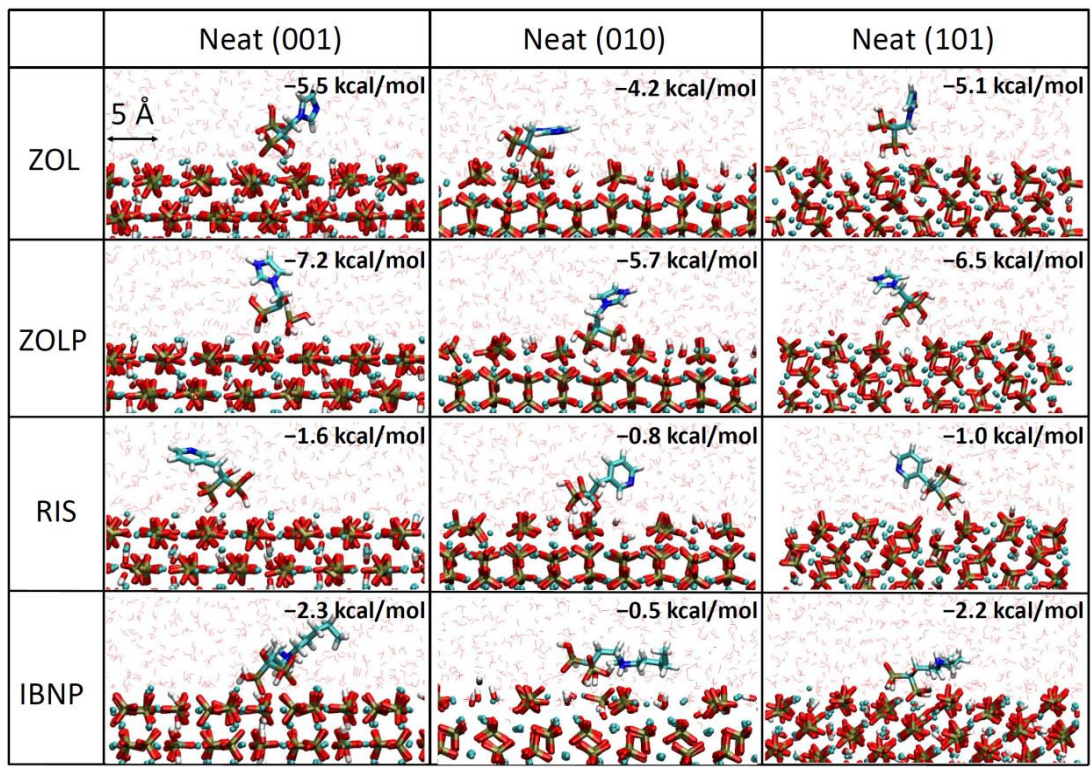

Figure 3. Snapshots that represent the binding modes and free energies of ZOL, ZOLP, RIS, and IBNP to neat HAP surfaces. ZOL, ZOLP, RIS, and IBNP bound to neat HAP surfaces primarily through their phosphonate parts. Water molecules are represented in the line style. The cyan, blue, red, white, green, and brown atoms represent carbon, nitrogen, oxygen, hydrogen, calcium, and phosphorous atoms, respectively.

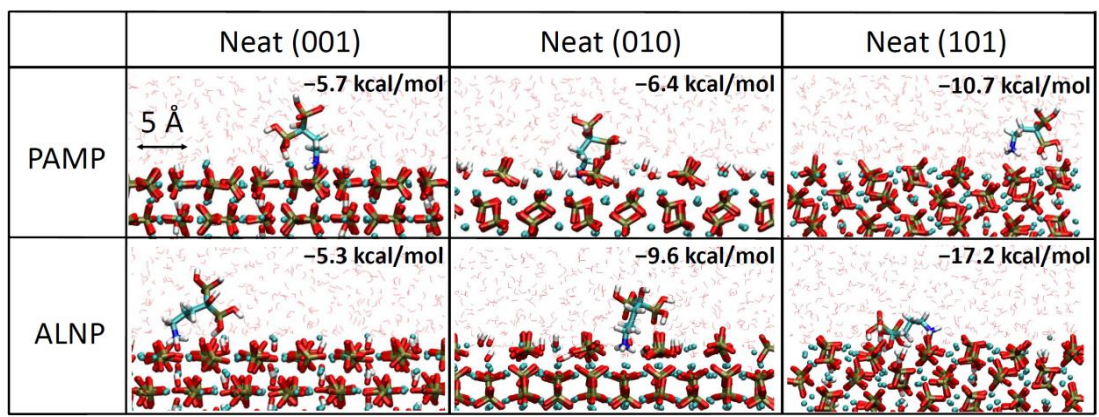

Figure 4. Snapshots that represent the binding modes and free energies of PAMP and ALNP to neat HAP surfaces. PAMP and ALNP bound to neat HAP surfaces primarily through their charged amine groups. At neat (010) surfaces, their charged amine groups were able to slip into the surface grooves. At neat (101) surfaces, both phosphonate and charged amine groups interacted with surface ions. Water molecules are represented in the line style. The cyan, blue, red, white, green, and brown atoms represent carbon, nitrogen, oxygen, hydrogen, calcium, and phosphorous atoms, respectively.

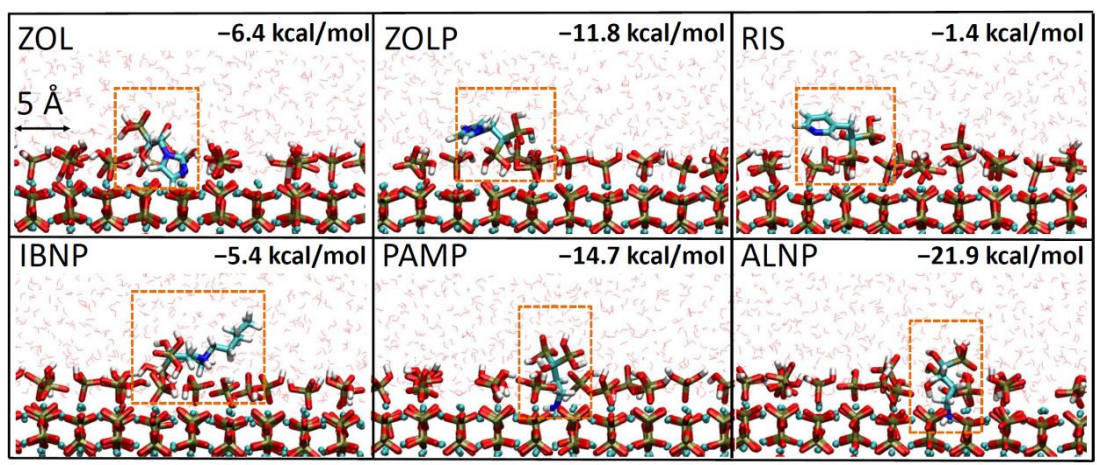

Figure 5. Snapshots that represent the binding modes and free energies of ZOL, ZOLP, RIS, IBNP, PAMP, and ALNP to (010)-pH7 surfaces. All six NBPs immersed in the surface grooves. Removing calcium and hydroxide ions at the surface by our protonation scheme enabled NBPs to penetrate surface grooves easily. Water molecules are represented in the line style, and NBPs are identified by a dashed rectangle. The cyan, blue, red, white, green, and brown atoms represent carbon, nitrogen, oxygen, hydrogen, calcium, and phosphorous atoms, respectively.

## Table of Contents

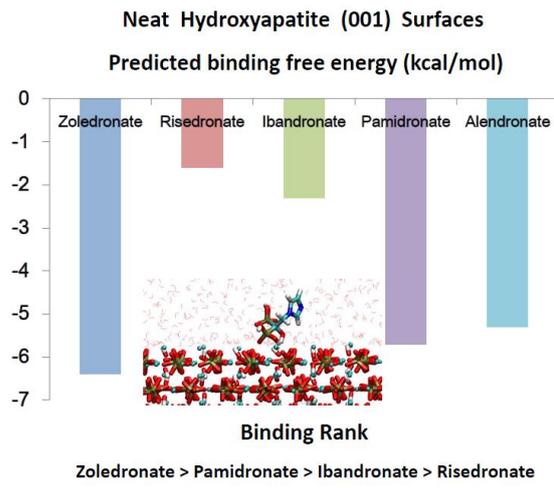

Neat Hydroxyapatite (001) Surfaces

Predicted binding free energy (kcal/mol)

Binding Rank

Zoledronate > Pamidronate > Ibandronate > Risedronate

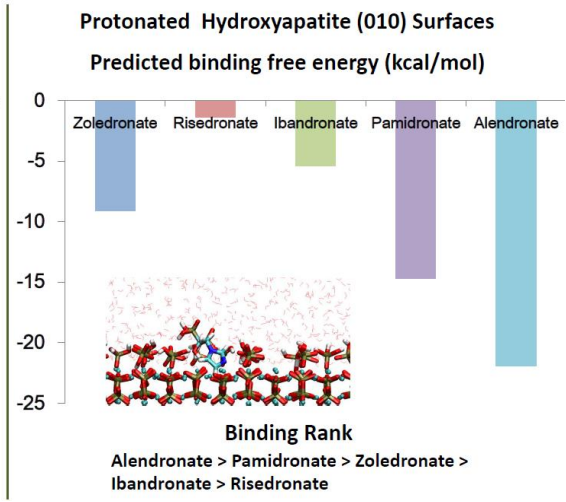

Protonated Hydroxyapatite (010) Surfaces

Predicted binding free energy (kcal/mol)

Binding Rank

Alendronate > Pamidronate > Zoledronate > Ibandronate > Risedronate

**Using Molecular Dynamics to Model the Binding Affinities of Nitrogen-Containing Bisphosphonates on Hydroxyapatite Surfaces**

Tzu-Jen Lin*

Department of Chemical Engineering, Chung Yuan Christian University, 200 Chung Pei Road, Chung Li District, Taoyuan City, 32023, Taiwan
* Corresponding author: tzujenlin999@gmail.com

Table A1. GAFF [1] atom types and atomic charges of protonated zoledronate (ZOLP).

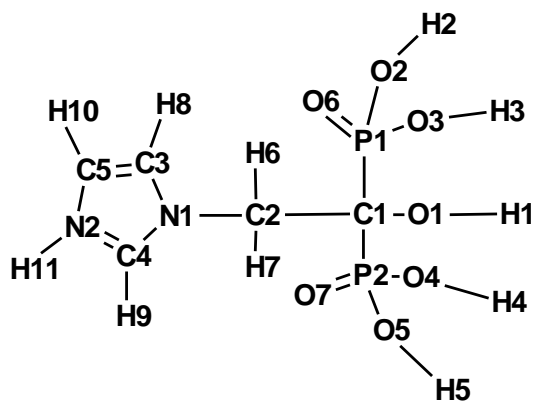

| Name | Type | Charge | Name | Type | Charge | Name | Type | Charge |
|------|------|--------|------|------|--------|------|------|--------|
| C1 | c3 | -0.497 | O4 | oh | -0.569 | N1 | na | -0.051 |
| P1 | p5 | 1.084 | H4 | ho | 0.443 | C3 | cc | -0.218 |
| P2 | p5 | 1.084 | O5 | oh | -0.569 | H8 | h4 | 0.266 |
| O1 | oh | -0.442 | H5 | ho | 0.443 | C4 | cc | 0.184 |
| H1 | ho | 0.451 | O6 | o | -0.617 | H9 | h5 | 0.194 |
| O2 | oh | -0.569 | O7 | o | -0.617 | C5 | cc | -0.051 |
| H2 | ho | 0.443 | C2 | c3 | 0.104 | H10 | h4 | 0.238 |
| O3 | oh | -0.569 | H6 | h1 | 0.119 | N2 | na | -0.236 |
| H3 | ho | 0.443 | H7 | h1 | 0.119 | H11 | hn | 0.390 |

Table A2. GAFF[1] atom types and atomic charges of neutral zoledronate (ZOL).

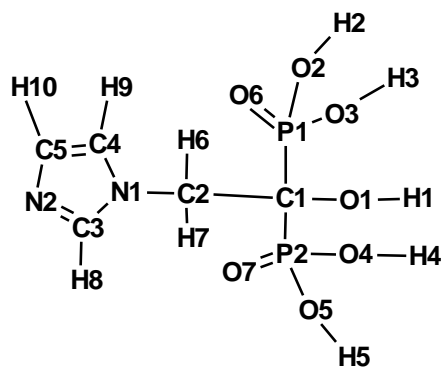

| Name | Type | Charge | Name | Type | Charge |
|------|------|--------|------|------|--------|
| C1 | c3 | -0.284 | O6 | o | -0.622 |
| P1 | p5 | 1.035 | O7 | o | -0.622 |
| P2 | p5 | 1.035 | C2 | c3 | 0.477 |
| O1 | oh | -0.526 | H6 | h1 | -0.022 |
| H1 | ho | 0.446 | H7 | h1 | -0.022 |
| O2 | oh | -0.586 | N1 | na | -0.232 |
| H2 | ho | 0.426 | C3 | cc | -0.506 |
| O3 | oh | -0.586 | H8 | h4 | 0.255 |
| H3 | ho | 0.426 | C4 | cc | 0.393 |
| O4 | oh | -0.586 | H9 | h5 | 0.061 |
| H4 | ho | 0.426 | C5 | cc | 0.315 |
| O5 | oh | -0.586 | H10 | h4 | 0.081 |
| H5 | ho | 0.426 | N2 | nc | -0.622 |

Table A3. GAFF[1] atom types and atomic charges of neutral risedronate (RIS).

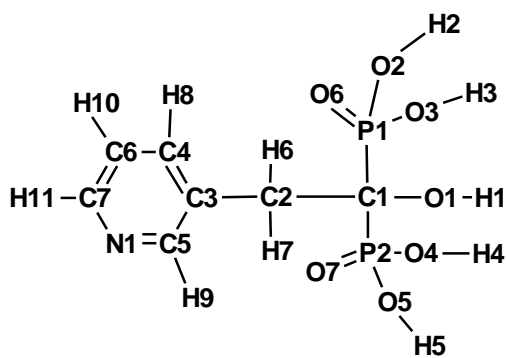

| Name | Type | Charge | Name | Type | Charge |
|------|------|--------|------|------|--------|
| C1 | c3 | -0.092 | O7 | o | -0.624 |
| P1 | p5 | 0.979 | C2 | c3 | 0.350 |
| P2 | p5 | 0.979 | H6 | hc | -0.035 |
| O1 | oh | -0.588 | H7 | hc | -0.035 |
| H1 | ho | 0.463 | C3 | ca | -0.519 |
| O2 | oh | -0.577 | C4 | ca | 0.191 |
| H2 | ho | 0.420 | H8 | ha | 0.100 |
| O3 | oh | -0.577 | C5 | ca | 0.594 |
| H3 | ho | 0.420 | H9 | h4 | -0.021 |
| O4 | oh | -0.577 | C6 | ca | -0.488 |
| H4 | ho | 0.420 | H10 | ha | 0.185 |
| O5 | oh | -0.579 | C7 | ca | 0.489 |
| H5 | ho | 0.420 | H11 | h4 | 0.035 |
| O6 | o | -0.624 | N1 | nb | -0.711 |

Table A4. GAFF [1] atom types and atomic charges of protonated pamidronate (PAMP).

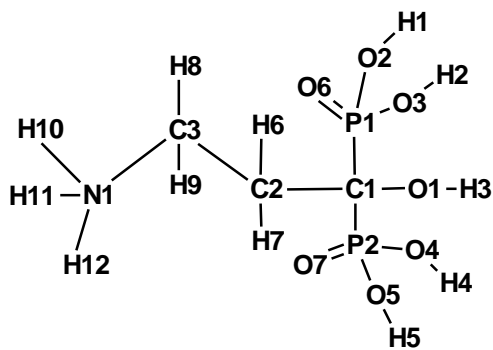

| Name | Type | Charge | Name | Type | Charge |
|------|------|--------|------|------|--------|
| C1 | c3 | -0.409 | O6 | o | -0.630 |
| P1 | p5 | 1.198 | O7 | o | -0.630 |
| P2 | p5 | 1.198 | C2 | c3 | -0.456 |
| O1 | oh | -0.491 | H6 | hc | 0.207 |
| H1 | ho | 0.430 | H7 | hc | 0.207 |
| O2 | oh | -0.605 | C3 | c3 | 0.184 |
| H2 | ho | 0.445 | H8 | hx | 0.123 |
| O3 | oh | -0.605 | H9 | hx | 0.123 |
| H3 | ho | 0.445 | N1 | n4 | -0.551 |
| O4 | oh | -0.605 | H10 | hn | 0.379 |
| H4 | ho | 0.445 | H11 | hn | 0.379 |
| O5 | oh | -0.605 | H12 | hn | 0.379 |
| H5 | ho | 0.445 | | | |

Table A5. GAFF [1] atom types and atomic charges of protonated alendronate (ALNP).

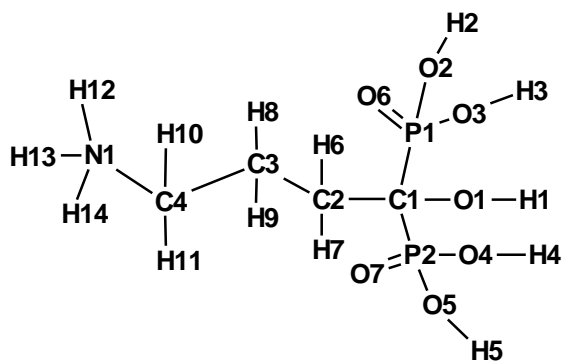

| Name | Type | Charge | Name | Type | Charge |
|------|------|--------|------|------|--------|
| C1 | c3 | 0.058 | O7 | o | -0.653 |
| P1 | p5 | 1.044 | C2 | c3 | -0.339 |
| P2 | p5 | 1.044 | H6 | hc | 0.135 |
| O1 | oh | -0.551 | H7 | hc | 0.135 |
| H1 | ho | 0.426 | C3 | c3 | 0.030 |
| O2 | oh | -0.606 | H8 | hc | -0.009 |
| H2 | ho | 0.456 | H9 | hc | -0.009 |
| O3 | oh | -0.606 | C4 | c3 | 0.396 |
| H3 | ho | 0.456 | H10 | hx | 0.006 |
| O4 | oh | -0.606 | H11 | hx | 0.006 |
| H4 | ho | 0.456 | N1 | n4 | -0.630 |
| O5 | oh | -0.606 | H12 | hn | 0.388 |
| H5 | ho | 0.456 | H13 | hn | 0.388 |
| O6 | o | -0.653 | H14 | hn | 0.388 |

Table A6. GAFF [1] atom types and atomic charges of protonated ibandronate (IBNP).

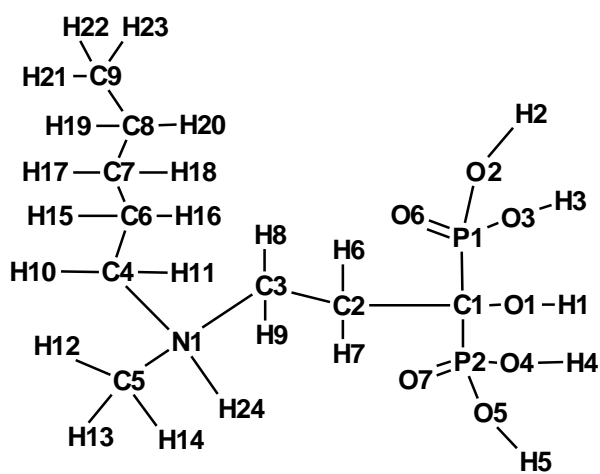

| Name | Type | Charge | Name | Type | Charge | Name | Type | Charge |
|------|------|--------|------|------|--------|------|------|--------|
| C1 | c3 | -0.097 | O7 | o | -0.632 | H10 | hx | 0.115 |
| P1 | p5 | 1.067 | C2 | c3 | -0.003 | H11 | hx | 0.115 |
| P2 | p5 | 1.067 | H6 | hc | 0.073 | C6 | c3 | -0.033 |
| O1 | oh | -0.519 | H7 | hc | 0.073 | H15 | hc | 0.026 |
| H1 | ho | 0.409 | C3 | c3 | -0.355 | H16 | hc | 0.026 |
| O2 | oh | -0.647 | H8 | hx | 0.149 | C7 | c3 | -0.030 |
| H2 | ho | 0.491 | H9 | hx | 0.149 | H17 | hc | 0.026 |
| O3 | oh | -0.647 | N1 | n4 | 0.348 | H18 | hc | 0.026 |
| H3 | ho | 0.491 | H24 | hn | 0.230 | C8 | c3 | 0.118 |
| O4 | oh | -0.647 | C5 | c3 | -0.569 | H19 | hc | 0.001 |
| H4 | ho | 0.491 | H12 | hx | 0.216 | H20 | hc | 0.001 |
| O5 | oh | -0.647 | H13 | hx | 0.216 | C9 | c3 | -0.266 |
| H5 | ho | 0.491 | H14 | hx | 0.216 | H21 | hc | 0.076 |
| O6 | o | -0.632 | C4 | c3 | -0.135 | H22 | hc | 0.076 |
|    |    |        |    |    |        | H23 | hc | 0.076 |

Figure A1. Atomic charge setups and numbers of protonated phosphates at HAP (001), (010), and (101) surface in a pH 7 condition. The atomic charge of phosphorus atom is 1.0, which is not showed in the figure because of tight spacing. The atomic charge setups for neat hydroxyapatite surfaces were demonstrated in previous work. [2]

| Surface | Atomic Charge Setup | | # of $H_2PO_4$ | # of $HPO_4$ | Surface Area* |
|---------|---------------------|--|----------------|--------------|---------------|
| (001)-pH7 |  |  | 34 | 14 | 32.6214*37.668 |
| (010)-pH7 |  |  | 14 | 6 | 37.6680*34.3750 |
| (101)-pH7 |  |  | 26 | 10 | 34.9787*32.6214 |

* The unit of length is Å, and the values were prior to simulations.

Figure A2. Free energy profiles of ZOL, ZOLP, RIS, IBNP, PAMP, and ALNP at neat HAP surfaces determined through umbrella sampling. The free energy reference point was NBPs located 15 Å from surfaces, and the positions of surface oxygen atoms were set as the zero point of reaction coordinates. Error bars are drawn as lines with filled colored areas.

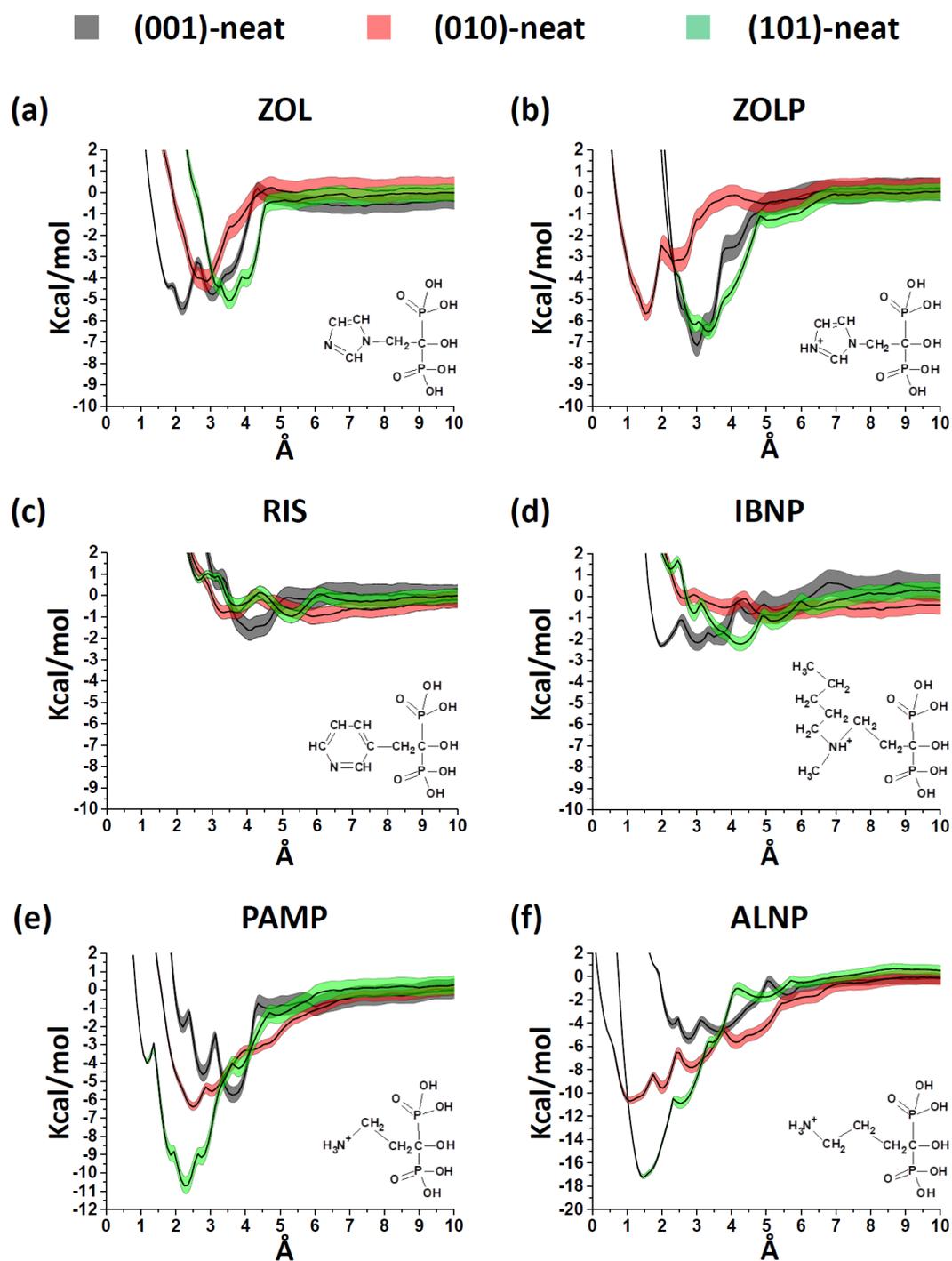

Figure A3. Free energy profiles of ZOL, ZOLP, RIS, IBNP, PAMP, and ALNP at protonated HAP surfaces in pH7 condition determined through umbrella sampling. The free energy reference point was NBPs located 15 Å from surfaces, and the positions of surface oxygen atoms were set as the zero point of reaction coordinates. The reaction coordinates below zero meant that the molecules immersed below surface oxygen atoms. Error bars are drawn as lines with filled colored areas.

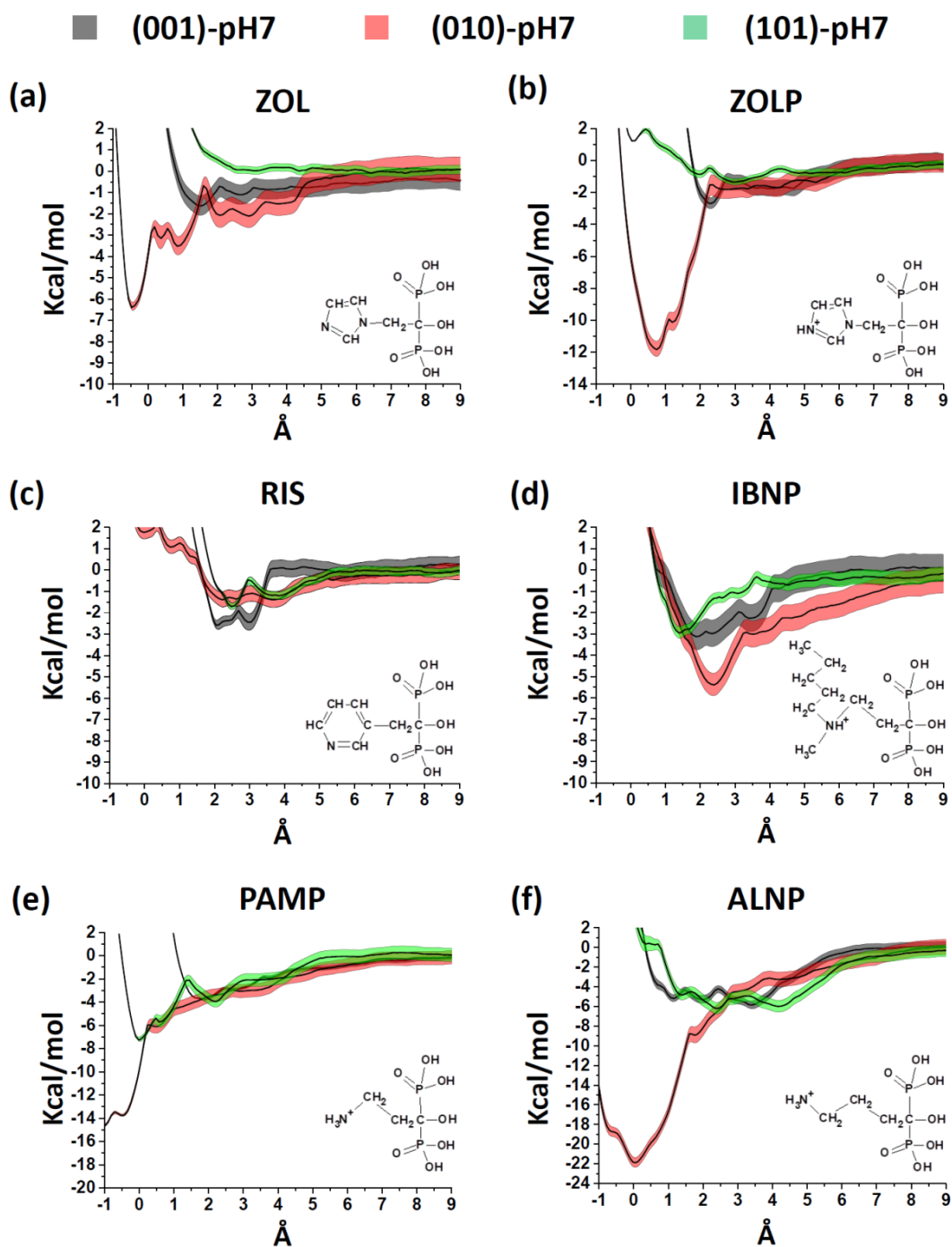

Table A7. Solvation free energies of ZOL, ZOLP, RIS, IBNP, PAMP, and ALNP by free energy perturbation. [3]

| NBP | Solvation Free Energy (kcal/mol) |
|------|------|
| ZOL | −27.0 |
| ZOLP | −59.6 |
| RIS | −23.9 |
| IBNP | −55.2 |
| PAMP | −67.6 |
| ALNP | −68.8 |